\documentclass[conference]{IEEEtran}
\IEEEoverridecommandlockouts
\usepackage{cite}
\usepackage{amsmath,amssymb,amsfonts}
\usepackage{graphicx}
\usepackage{textcomp}
\usepackage{mathtools}
\usepackage{amsmath}
\usepackage{algorithm}

\usepackage{url}
\usepackage{hyperref}
\usepackage{caption}
\usepackage{blindtext}
\usepackage{enumitem}
\usepackage{varwidth}
\usepackage[noend]{algpseudocode}
\usepackage{xcolor}
\def\BibTeX{{\rm B\kern-.05em{\sc i\kern-.025em b}\kern-.08em
    T\kern-.1667em\lower.7ex\hbox{E}\kern-.125emX}}
\begin{document}

\title{Motif Identification using CNN-based Pairwise Subsequence Alignment Score Prediction\\
\thanks{Identify applicable funding agency here. If none, delete this.}
}

\author{\IEEEauthorblockN{Ethan Moyer}
\IEEEauthorblockA{\textit{School of Biomedical Engineering, Science and Health Systems} \\
\textit{Drexel University}\\
Philadelphia, PA \\
https://orcid.org/0000-0002-8023-3810}
\and
\IEEEauthorblockN{Anup Das}
\IEEEauthorblockA{\textit{College of Engineering} \\
\textit{Drexel University}\\
Philadelphia, PA \\
https://orcid.org/0000-0002-5673-2636}

}

\maketitle

\begin{abstract}
A common problem in bioinformatics is related to identifying gene regulatory regions marked by relatively high frequencies of motifs, or deoxyribonucleic acid sequences that often code for transcription and enhancer proteins. Predicting alignment scores between subsequence k-mers and a given motif enables the identification of candidate regulatory regions in a gene, which correspond to the transcription of these proteins. We propose a one-dimensional (1-D) Convolution Neural Network trained on k-mer formatted sequences interspaced with the given motif pattern to predict pairwise alignment scores between the consensus motif and subsequence k-mers. Our model consists of fifteen layers with three rounds of a one-dimensional convolution layer, a batch normalization layer, a dense layer, and a 1-D maximum pooling layer. We train the model using mean squared error loss on four different data sets each with a different motif pattern randomly inserted in DNA sequences: the first three data sets have zero, one, and two mutations applied on each inserted motif, and the fourth data set represents the inserted motif as a position-specific probability matrix. We use a novel proposed metric in order to evaluate the model's performance, $S_{\alpha}$, which is based on the Jaccard Index. We use 10-fold cross validation to evaluate out model. Using $S_{\alpha}$, we measure the accuracy of the model by identifying the 15 highest-scoring 15-mer indices of the predicted scores that agree with that of the actual scores within a selected $\alpha$ region. For the best performing data set, our results indicate on average 99.3\% of the top 15 motifs were identified correctly within a one base pair stride ($\alpha = 1$) in the out of sample data. To the best of our knowledge, this is a novel approach that illustrates how data formatted in an intelligent way can be extrapolated using machine learning.
\end{abstract}


\section{Introduction}
\label{sec:intro}

\begin{figure}[htp]
    \centering
    \includegraphics[width=0.8\linewidth]{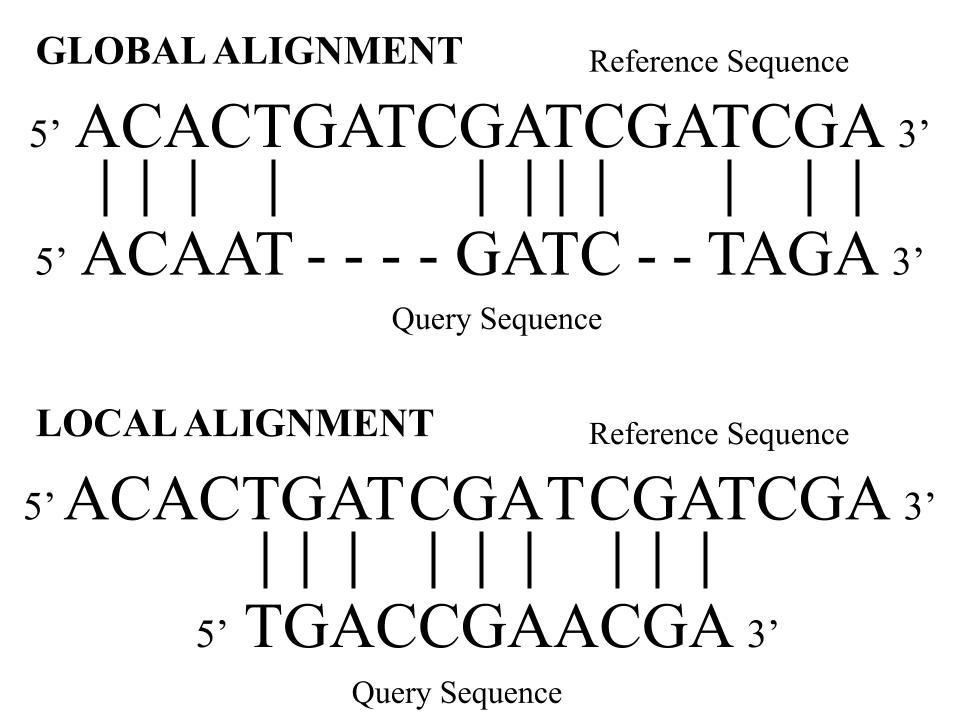}\caption{Local vs. Global Alignment. In general, DNA is composed of a permutation of the four nucleotides [adenine (A), thymine (T), cytosine (C), guanine (G)] and an ambiguous base (N).}
    \label{figdnaalign}
\end{figure}

\IEEEPARstart{M}{easuring} the similarity of two sequences is a well known problem called sequence alignment. This topic includes a vast category of methods for identifying regions of high similarity in biological sequences, such as those in deoxyribonucleic Acid (DNA), ribonucleic acid (RNA), and protein \cite{Haque2009pairwise}. Specifically, DNA pairwise sequence alignment (PSA) methods are concerned with finding the best arrangement of two DNA sequences. Some historically notable dynamic programming PSA methods are the Needleman-Wunsch (NW) algorithm for global alignment \cite{spr1970general} and Smith-Waterman (SW) algorithm for local alignment \cite{smith1981identification}. The main difference between global and local alignment is related to the difference in length of the two sequences: global alignment attempts to find the highest-scoring end-to-end alignment between two sequences of approximately the same length, and local alignment searches for local regions of high similarity between two sequences with different lengths \cite{ebi2020pairwise}. \autoref{figdnaalign} shows this difference between local and global DNA alignment with two sequences aligned in a 5' (i.e. five prime) to 3' direction. In molecular biology, this orientation refers to the directionality of the carbon backbone in DNA. The top subfigure displays global alignment where a query sequence is aligned end-to-end with a reference. The bottom subfigure displays local alignment where a short query sequence is most optimally aligned with a longer reference sequence. This latter alignment displays how the query sequence is approximately equal to a subsequence of the reference sequence. 


In this way, local alignment methods recognize approximate subsequence matches of a query sequence with respect to a given reference sequence. One common paradigm utilizing local alignment is to examine similarities between a query sequence and specific k-long subsequences in a given gene, known as k-mers, found within the reference sequence. Traditional local alignment algorithms calculate these scores between the query sequence and each k-mer in the reference sequence. The aim of this research is to identify where the most likely subsequence matches of the query sequence occur in each reference sequence using machine learning methods. One such type of query sequence that is of high biological significance is a sequence motif, which are short reoccurring subsequences of DNA \cite{d2006dna}. Therefore, this research follows the ability of machine learning methods to gauge the relative enrichment of various representations of motifs (or motif patterns) in independent reference sequences. More specifically, the efficacy of identifying motif enrichment in sequences is explored using a one-dimensional (1-D) convolution neural network (CNN).

Four different data sets are generated, each with a different motif pattern randomly inserted in approximately 10,000 reference sequences: the first three data sets have zero, one, and two mutations applied on each inserted motif, and the fourth data set represents the inserted motif as a position-specific probability matrix (PPM). In this data structure, each nucleotide position corresponds to a frequency of nucleotides~\cite{wang2020motto}. These distinct motif patterns help display how the CNN model can recognize both subsequence matches with exact, inexact, and probabilistic motifs. Each sample in a given data set consists of artificial sequences enriched with a given motif pattern at a frequency between five and fifteen occurrences per 1,000 base pairs (bp). These samples are split into 986 overlapping 15-mers with a corresponding calculated local alignment score from the BioPython Aligner \cite{cock2009biopython}. These sores are then predicted using a CNN with 10-fold cross validation. In order to measure the performance of the model, the average out of sample mean squared error (MSE), R2, and accuracy scores are reported.

While the MSE of the model trained on each data set is not representative of the model's effectiveness, the Jaccard Index and $S_{\alpha}$, a novel modified version of the Jaccard Index, are better suited to capture accuracy of the model. The standard MSE is not suitable for this problem because it inherently only displays differences between predicted and actual values. Since our aim is to locate those highest-scoring 15-mers, we need a metric that determines at which positions they occur and with what accuracy (see \autoref{cnnevaul}). \color{black} This new metric, $S_{\alpha}$, measures the degree of similarity between two sets where each pair of elements can be different by at most $\alpha$. Because of the plateauing nature of this metric as seen in each data set and the risks involved in increasing alpha, only $S_{0}$ to $S_{5}$ are reported.

In implementing this new metric, the accuracy of the model increases dramatically across all four data sets compared to the Jaccard Index. This indicates that while the model is not able to precisely identify the highest-scoring k-mers exactly, it is able to accurately identify their local region. As expected, the model's accuracy is far higher for the data sets with relatively simple inserted motif patterns--non-probabilistic consensus motifs--compared to that of the data set with more complex inserted motif patterns, such as consensus PPM. 

\section{Background}

Clusters of motifs across a genome strongly correlate to a gene regulatory regions \cite{frith2003cluster}. These regions are especially important for motif enrichment analysis, where known motifs are identified in the regulatory sequence of a gene in order to determine which proteins (transcription factors and enhancers) control its transcription~\cite{mcleay2010motif} \cite{lesluyes2014differential}. Motif enrichment analysis is only relevant given that the regulatory region of a gene is known, otherwise the sequence under study may be from a non-coding region of an organism's genome or an untranslated region of a gene \cite{liu2000bioprospector}. Given that the regulatory region of a gene is unknown, one frequently used approach to identifying it is to first locate sequences enriched with highly conserved motifs. Fortunately, many motifs that have been discovered are common amongst genes serving a similar role across organisms, such as a negative regulatory region for eukaryotes \cite{iniguez2000common}. Finding these conserved motifs may facilitate the identification of the  regulatory regions in a gene. For that reason, identifying the exact or relative positions of a given motif in a gene or sequence is a relevant inquiry in the process for classifying candidate regulatory regions of a gene. 

A software toolkit known as MEME Suit includes three different methods for motif-sequence searching \cite{Bailey2009aa}: FIMO (Find Individual Motif Occurrences) \cite{Grant2011aa}, GLAM2SCAN (Gapped Local Alignment of Motifs SCAN) \cite{Frith2008aa}, and MAST (Motif Alignment and Search Tool) \cite{Bailey1998aa}.

FIMO focuses on scanning both DNA and protein sequences for a given motif represented as PPM. This software tool calculates the log-likelihood ratio score, p-value, and q-value (false discovery rate) for each subsequence position in a sequence database \cite{Grant2011aa}. 

Typically, GLAM2SCAN performs a Waterman-Eggert local alignment between motifs found by GLAM2, its companion motif-finding algorithm, and a sequence database. These local alignment scores are generated from an aligner programmed with position specific residue scores, deletion scores, and insertion scores returned from the GLAM2 algorithm. The $n$ highest alignments are returned to the user \cite{Frith2008aa}. 

MAST locates the highest-scoring $n$ subsequences with respect to a motif described as a position-specific score matrix. Using the QFAST algorithm, MAST calculates the p-value of a group of motif matches. This is accomplished by first finding the p-value of each match (position p-value') and normalizing it for the length of the motif ('sequence p-value'). Then each of these normalized p-values are multiplied together to find the statistical significance across all located motifs in the database ('combined p-value') \cite{Bailey1998aa}. 

\section{Data Analysis \& Curation}

A single data set contains approximately 10,000 randomly generated DNA sequences, each 1,000 bp long. The number of samples vary slightly from one to another due to some inconsistencies that are removed in prepossessing. A 15-mer motif is inserted into each sample anywhere from five to fifteen times. Four separate data sets of this structure are created where a different motif pattern is inserted randomly into each sequence. The  first  three  data  sets  have  zero, one,  and  two  mutations  applied  on  each  inserted  motif. These mutations are applied in order to determine whether the proposed model has the potential to identify consensus motifs and non-exact consensus motifs across many sequences. Since motifs mostly exist as profiles where each base pair position corresponds to a frequency table of nucleotides, the fourth data set is created where the inserted motifs are based off of a PPM \cite{das2007survey}.

\autoref{motifequ} is used to calculate the PPM indicated by matrix $M$ given a set of candidate motifs, or sequences that are thought to be from the same motif PPM. This equation counts the number of occurrences of each nucleotide in set $\gamma$ for each nucleotide position across all motifs, where $\gamma = \{A, T, C, G\}$; $I =  \{0,1\}$ represents an indicator function, where $I(x=\gamma)$ is 1 if $x=\gamma$ and 0 otherwise; $i {\displaystyle \in }$ (1, ..., L), where L is the length of each motif; and $j {\displaystyle \in }  (1, ..., N)$, where N is the number of motifs.

    \begin{equation}\label{motifequ}
       M_{\alpha,k} = \frac{1}{N}\sum^{N}_{i=1}I(X_{i,j} = \gamma)
    \end{equation}

In order to apply \autoref{motifequ} on candidate motifs, the DNA sequence data must be formatted as nucleotide position counts shown in \autoref{dnatoppm}. This figure illustrates the conversion of a list of candidate motifs to matrix $M_{counts}$ and then to $PPM$ using \autoref{motifequ}. 
While \autoref{dnatoppm} displays this process for five 10-mers, the fourth data sets in this work relies on profiles built from ten 15-mers.

\begin{figure}[H]

\begin{center}
\begin{varwidth}{\textwidth}
\centering
\begin{description}
\item TACAGAGTTG
\item CCATAGGCGT
\item TGAACGCTAC
\item ACGGACGATA
\item CGAATTTACG
\end{description}

$\downarrow$
\end{varwidth}
\end{center}
\centering
$M_{counts}$ =
\begin{tabular}{l|l|l|l|l|l|l|l|l|l|l|}
\cline{2-11}
\textbf{A} & 1 & 1 & 3 & 3 & 2 & 1 & 0 & 2 & 1 & 1 \\ \cline{2-11} 
\textbf{T} & 2 & 0 & 0 & 1 & 1 & 1 & 1 & 2 & 2 & 1 \\ \cline{2-11} 
\textbf{C} & 2 & 2 & 1 & 0 & 1 & 1 & 1 & 1 & 1 & 1 \\ \cline{2-11} 
\textbf{G} & 0 & 2 & 1 & 1 & 1 & 2 & 3 & 0 & 1 & 2 \\ \cline{2-11} 
\end{tabular}
\vspace{5px}
\newline
\centering
$\downarrow$

\resizebox{0.8\textwidth}{!}{\begin{minipage}{\textwidth}
\vspace{8px}
$PPM$ =
\begin{tabular}{l|l|l|l|l|l|l|l|l|l|l|}
\cline{2-11}
\textbf{A} & 0.2 & 0.2 & 0.6 & 0.6 & 0.4 & 0.2 & 0.0 & 0.4 & 0.2 & 0.2 \\ \cline{2-11} 
\textbf{T} & 0.4 & 0.0 & 0.0 & 0.2 & 0.2 & 0.2 & 0.2 & 0.4 & 0.4 & 0.2 \\ \cline{2-11} 
\textbf{C} & 0.4 & 0.4 & 0.2 & 0.0 & 0.2 & 0.2 & 0.2 & 0.2 & 0.2 & 0.2 \\ \cline{2-11} 
\textbf{G} & 0.0 & 0.4 & 0.2 & 0.2 & 0.2 & 0.4 & 0.6 & 0.0 & 0.2 & 0.4 \\ \cline{2-11} 
\end{tabular}
\end{minipage}}
\caption{The conversion of five candidate subsequence motifs to PPM using Equation \ref{motifequ}.}
\label{dnatoppm}

\end{figure}

\section{Feature \& Output Selection}

\begin{figure}[H]
    \centering
    \includegraphics[width=0.8\linewidth]{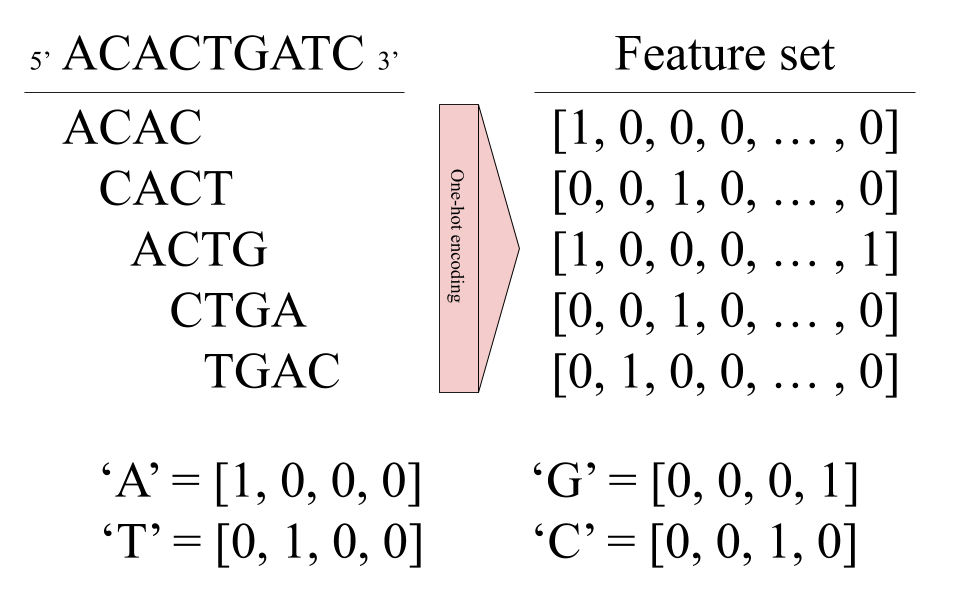}\caption{DNA subsequence k-mer formatting by one-hot encoding nucleotides.}
    \label{dnaformatting}
\end{figure}

In order to format the sequence data into a structure that is both recognizable and meaningful to a CNN, we first split each sequence into a list of overlapping 15-mers. Next, we generate a one-hot encoding for each nucleotide in the 15-mers. The resulting feature set is composed of 60 values. \autoref{dnaformatting} displays this process using a small subsequence example formatted as 4-mers.

To obtain the target values, each of these 15-mers are pairwise aligned with the consensus motif for the given data set motif pattern using the SW algorithm. Given two sequences, $a$ of length $n$ and $b$ of length $m$, this algorithm begins by defining an $n + 1$ by $m + 1$ matrix $H$. The first column and first row are assigned $0$, and the following recurrence relation is applied to assign the rest of the values in $H$.

\[ H(i, j) = max
  \begin{cases}
    H(i - 1, j - 1) + \sigma(a_{i}, b_{j})\\
    H(i, j - 1) + W\\
    H(i - 1, j) + W\\
    0
  \end{cases}
\]

where W is a gap score and $\sigma$ is a score matrix such that

\[ \sigma(a_{i}, b_{j}) =
  \begin{cases}
    +1 & \quad \text{if } a_{i} = b_{j}\\
    -2 & \quad \text{if } a_{i} \neq b_{j}
  \end{cases}
\]
In the case when $a_{i} = b_{j}$, $\sigma$ returns a match score of $+1$, and in the case when $a_{i} \neq b_{j}$, $\sigma$ returns a mismatch score of $-2$. The gap score, $W$, is assigned $-2.5$. The match, mismatch, and gap score can be configured for different alignments. These parameters are used because they are the most optimal for this type of local alignment \cite{al2016survey}. Once $H$ is assigned its values, the best alignment is obtained by finding the maximum value in $H$ and tracing back the matrix elements that led up to this maximum. In this way, the maximum value in $H$ defines the optimal path in $H$ for the best alignment between sequences $a$ and $b$ \cite{smith1981identification}. \color{black} The calculated alignment scores are normalized based on the maximum alignment score in each sample.
\section{Methods} \label{methods}

\subsection{CNN Model Evaluation} \label{cnnevaul}

Although the MSE loss function is  effective at penalizing large differences between predicted and target values, such as outliers in the data, it does not successfully represent the predictive power of the model given the scope of the problem \cite{qi2014robust}. In the data, the target value from each sample ranges from zero to one. This range already generates an inherently small MSE. Even when the MSE for each sample is normalized, the metric is overshadowed by the overwhelming majority of the predicted values that were approximately equal to the global mean of each sample. In other words, the MSE as a metric does not capture the correct information pertaining to the five to fifteen inserted motif patterns in each sample due to a large unequal distribution of such scores that deviate from the global mean. This problem is analogous to that of an unequal class distribution in a classification problem.

The goal of the model is to score the CNN based on its ability to locate the 15 highest-scoring 15-mers, because we inserted a motif pattern at most 15 times into a single sample. Since this network deals with continuous values instead of discrete classes, initially we cannot be certain of the 15-mer to which a 15-mer score at any index $i$ corresponds. However, a higher scoring 15-mer has a greater probability of corresponding to that of a motif, whereas the lower scoring 15-mers carry little information. This is due to the fact that each score in the data is generated from a local alignment between 15-mer and the given consensus motif. In this way, only the highest 15-scoring 15-mers are of interest. As previously mentioned, we indicate that there is an unequal distribution between the number of scores corresponding to that of each inserted motif and the global mean of each sample. Using these observations, we rationalize that we only have to examine the 15 highest-scoring indices. This generality that the 15 highest-scoring idicies correspond to the inserted motif patterns is further supported by the notion that probability of observing a random 15-mer exactly equal or similar to the inserted motifs is relatively low.

Thus, the indices of the predicted 15 highest-scoring 15-mer inherently hold information about the position of possible inserted motif patterns because it is at these indices at which the local alignment is conducted. Due to the low likelihood of observing a false positive (when a 15-mer is identified as a motif but in all actuality is not one), we create a one-to-one correspondence between the indices of the actual motif indices and that of the predicted motifs using high local alignment scores. The accuracy of this one-to-one correspondence can be measured using the Jaccard Index given in \autoref{jarrardindexequ}.

\begin{figure}[H]
    \centering
        \begin{equation}\label{jarrardindexequ}
        J(A, B) = \frac{|A \cap B|}{|A \cup B|}
        \end{equation}
\end{figure}

We propose a more generalized index, $S_{\alpha}$, in \autoref{jarrardindexequadp} which measures the similarity of two sets with an allowed margin of error of $\alpha$. Because of the high locality of local alignment score predictions and due to the fact that the highest-scoring 15-mers can still be found from examining the immediate region of a prediction, this margin of error serves as a heuristic for motif identification. In this metric, two items are considered identical if they are no more than $\alpha$ away from each other. In the scope of this work, sets $A$ and $B$ contain the indices of the 15 highest-scoring 15-mers of the actual data and predicted data, respectively. When $\alpha = 0$, $S_{0}(A, B)$ in \autoref{jarrardindexequ} is identical to $J(A, B)$ in \autoref{jarrardindexequadp}. Conversely, as $\alpha$ increases, the allowed distance between indices in sets $A$ and $B$ increases. For example, when $\alpha = 2$, a predicted 15-mer index $i$ and actual 15-mer index $i + 2$ are considered the same. 

\begin{figure}[H]
    \centering
        \begin{equation}\label{jarrardindexequadp}
        J(A, B \mid \alpha) = S_{\alpha}(A, B) = \frac{|\bigcup\limits_{ \mu = 0
        }^{\alpha} A\cap \{ x + \mu \mid x \in B \}|}{|A \cup B|} 
        \end{equation}
\end{figure}

The following process is an algorithm to calculate a modified version of the Jaccard Index. Using the $argsort$ function in NumPy, we examine the indices that order both the actual outputs and the predicted outputs. In looping through the each of the top $n$ indices of the predicted outputs, we count the number of them which are contained in the list of indices of the actual outputs. The process returns the score as count over the maximum possible value, which in this case is $n$. This is implemented in Algorithm \autoref{cnnscorestride}

\begin{algorithm}[H]
\caption{Measuring Jaccard Index with stride $\alpha$}\label{cnnscorestride}
\begin{algorithmic}[1]
\Procedure{$s_{\alpha}$}{}
\State $\textit{n} \gets \text{number of highest-scoring k-mers to analyze}$
\State $\textit{score} \gets 0$
\State $\textit{act\_outputs} \gets \text{actual outputs}$
\State $\textit{pred\_outputs} \gets \text{outputs from CNN}$
\State $\textit{act\_indxs} \gets \text{indices that would sort }\textit{act\_outputs}$
\State $\textit{pred\_indxs} \gets \text{indices that would sort } \textit{pred\_outputs}$ 
\State \emph{outerloop}:
\For{$i$ := 1 to $n$} 
\State $\textit{pred\_indx} \gets \textit{pred\_indxs(i)}$.
\For{$j$ := 0 to $\alpha$} 
\If {$\textit{pred\_indxs} \in \textit{act\_indxs} - j$}
\State $score \gets score+1$.
\State \textbf{goto} \emph{outerloop}.
\EndIf
\If {$\textit{pred\_indxs} \in \textit{act\_indxs} + j$}
\State $score \gets score+1$.
\State \textbf{goto} \emph{outerloop}.
\EndIf
\EndFor
\EndFor
\State $normalized\_score \gets score / n$.
\EndProcedure
\end{algorithmic}
\end{algorithm}

\section{Results}

Each of the four data sets is characterized by 10,000 samples where each sample contains a sequence that is 1,000 bp in length. In each sample, a motif pattern is inserted randomly anywhere from five to fifteen times. The first three data sets include inserted motif patterns with zero, one, and two mutations. The fourth data set includes an inserted motif pattern represented based on a PPM. Each data set is evaluated using out of sample data generated from 10-fold cross validation based on eight metrics: MSE, R2, and $S_{0}$-$S_{5}$.

\begin{table}[H]
    \centering
    \includegraphics[width=1.0\linewidth]{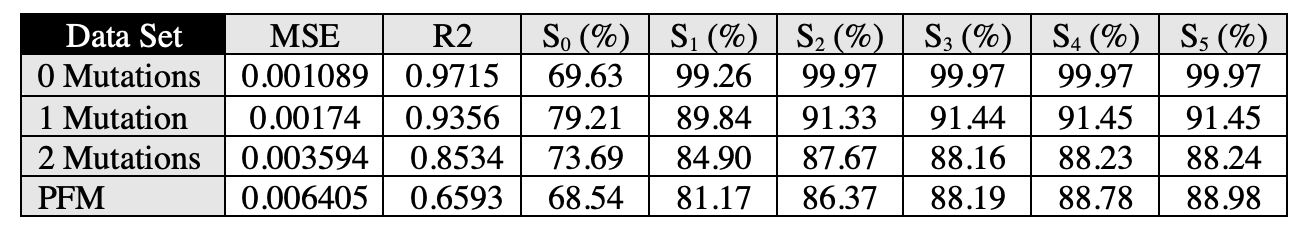}\caption{CNN Results. The average out of sample MSE, R2, and $S_{0}$-$S_{5}$ for each data set.}
    \label{cnnresults}
\end{table}

A fifth analysis is conducted with another data set using a motif representation similar to that of the fourth data set with the MafK transcription factor from the BATCH1 regulatory gene \cite{jaspar2019}. This motif is a 15-mer with a less conserved consensus sequence compared to that of the former four data sets. While this data set did not perform as well as the other four data sets with a $S_{9}$ of 45.3\%, this analysis brought to light the consideration of the aligner scoring matrix as another hyperparameter to this work. 

As it turns out, the performance of the model varies greatly with the chosen match score, mismatch score penalty, and gap score penalty for the currently implemented alignment method. For instance, the $S_{9}$ varies from 33.7\% to 52.6\% with different scoring hyperparameters. The former result is derived from an aligner with a match score of +2.0, mismatch score penalty of -3.0, and gap score penalty of -3.5, whereas the latter result is derived from an aligner with a match score of +2.0, mismatch score penalty of -4.0, and gap score penalty of -4.5. It is currently unclear what aligner hyperparameters are most optimal for this more complex data set and the original four data sets explored in the work. Although there is evidence to suggest that aligner scoring matrices vary with the type of inserted motif pattern, it is unclear whether the most optimal hyperparameters change from motif to motif. 

One possible interpretation of the dependence of the model's chosen evaluation metric, $S_{\alpha}$, on the aligner hyperparameters is related to the fact that the CNN predicts alignment scores that are normalized within each sample. Therefore, the farther these highest-scoring scores are from the global mean, the more likely that the proposed metric will be able to recognize inserted motifs. Conversely, when analyzing a data set with a less conserved motif consensus sequence, such as that of the MafK transcription factor, the alignment scores are closer to the global mean of each sample. This in turn makes recognizing the indices of the highest-scoring segments more challenging. It follows that the aligner hyperparameters which capitalize on increasing this difference are most favorable for all motifs, regardless of pattern.

\subsection{Convolution Neural Network (CNN) Architecture}

CNN is a class of deep learning models which can infer patterns based on data formatted as a grid structure, such as a set of prices over time for stock or a grid representation of pixels in an image (add reference for these architectures). These Artificial Neural Netowrk (ANNs) use a linear mathematical operation called convolution in at least one of their layers~\cite{bengio2020deep}. The convolution operation is commonly identified by the following two equations:

\begin{figure}[H]
    \centering
    \begin{equation}\label{convintegral}
        s(t) = \int x(a)w(t - a)da
    \end{equation}
    \begin{equation}\label{convfunction}
        s(t) = (x*w)(t)
    \end{equation}
    \label{figconvequ}
\end{figure}

\begin{figure}[H]
    \centering
    \includegraphics[width=1\linewidth]{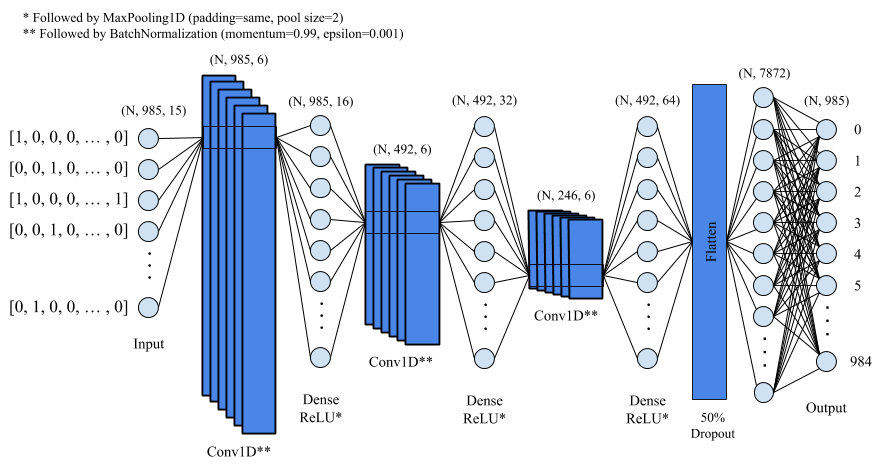}
    
    \caption{CNN model. (create better caption)}
    
    \label{figmodel}
\end{figure}

\autoref{convintegral} explicitly denotes the equation for convolution, whereas \autoref{convfunction} displays how an asterisk can be used to for the linear operation. In both equations, $x$ is referred to as the input. Typically, this is formatted as a multidimensional array, or a tensor, that matches the size and dimensions of the data. The second argument is $w$, representing a kernel, which stores parameters for the model also formatted as a tensor. This argument is adapted throughout the training process of the model. The output of both functions, $s$, is called the feature map of the convolution layer. This is what is fed into the next layer of the network \cite{bengio2020deep}. Hidden layers are generated from applying a kernel, or filter, of weights over the receptive field of the inputs. More specifically, the hidden layer is computed based off of the filter weights and the input layer as it strides across the feature space \cite{ref19}. This operation can either compress or expand input space depending on the applied kernel \cite{ref18}. This paradigm is followed by rounds of activations, normalizations, and pooling \cite{ref18}. The model typically ends with a fully connected layer to compute its outputs \cite{ref19}. The proposed model is represented in \autoref{figmodel} [cite my paper].

The model is marked by three rounds of a 1-D convolution layer, a batch normalization layer, a dense layer, and a 1-D maximum pooling layer. After these 12 layers, the model finishes off with a 50\% dropout layer, a flattened layer, and finally a fully connected layer corresponding to the 986 alignment scores for each sample \cite{schmidhuber2015deep} \cite{lecun2015deep}.

The model described above is ran on all four data sets for 100 epochs with a batch size of 80 and compiled with the Adam optimizer (learning rate=0.001, beta 1=0.9, beta 2=0.999, epsilon=1e-07). Of the 10,000 samples in each dataset, 80\% is reserved for training the network and the remaining 20\% is used for validation after each epoch. 
For its loss function, the model relies on Mean Squared Error (MSE), which is calculated between predicted values ($y_{pred}$) and target values ($y_{act}$) with the following formula in Equation \ref{mse}:
\begin{figure}[H]
    \centering
        \begin{equation}\label{mse}
        MSE(y_{pred}, y_{act}) = \frac{1}{n}\sum_{i = 1}^{n} (y_{pred,i} - y_{act,i})
        \end{equation}
    \label{fig:my_label}
\end{figure}

\section{Discussion}
As displayed in this work, deep learning models, such as a CNN, have the capacity to recognize and predict the positions of an inserted motif with great accuracy. Furthermore, data structures can be devised to take advantage of unequal class distributions in regression problems as highlighted by the design of k-mer data representation in this work and the incorporation of $S_{\alpha}$ as a novel evaluation metric.

In analyzing the results in \autoref{cnnresults}, there is a characteristic pattern between the accuracy metrics across each data set. For instance, in comparing $S_{0}$-$S_{5}$ for the first data set with zero mutations applied on each inserted motif, the score monotonically increases with an increasing $\alpha$. This is evident for the three other data sets as well. With respect to this particular trend, it is expected that as $\alpha$ increases, the score will also increase since $\alpha$ relates directly to the allowed margin of error, making $S_{\alpha}$ less conservative. 

Additionally, the model's accuracy is far higher for the data sets with relatively simple inserted motif patterns, such as nonmutated and mutated consensus motifs, compared to that of the fourth data set with a PPM motif pattern. This relationship can be explained by the process by which the scores for each 15-mer are calculated. For a given 15-mer, a score is computed based on its local alignment with a given consensus motif. For the first data set, these local alignment scores generated are derived from each inserted motif, whereas in the latter three data sets, the scores are not necessarily derived from each data set's consensus motif since the motif patterns support variable inserted motif.

In all data sets, the largest increase in $S_{\alpha}$ appears to be between the $S_{0}$ and $S_{1}$. After this point, change in $S_{\alpha}$ plateaus after a given $\alpha$. With the consideration that the likelihood of observing a false positive is relatively low, this indicates that the addition of stride $\alpha$ is well-advised. This is the case because the increase in $\alpha$ only influences $S_\alpha$ up to a certain point. It is expected that as $\alpha \xrightarrow{} \beta$, where $\beta$ is the maximum $\alpha$ on either side of a given motif index, $S_{\alpha}\xrightarrow{} 1$ because every single $n$ indices will be covered by the stride ${\alpha}$. In the case that $S_{\alpha} \xrightarrow{} 1$, the certainty for each identified motif decreases with increasing $S_{\alpha}$ regardless; however, the absence of this limit in the data indicates that the certainty of the identified motifs does not decreases dramatically from $S_{0}$ to $S_{5}$. Furthermore, the presence of a plateauing $S_{\alpha}$ supports the thought that a decrease in the certainty of an identified motif is negligible. This analysis can be drawn further in noticing that the point at which $S_{\alpha}$ plateaus increases as the complexity of the motif pattern increases. In the case of a more complex motif pattern, such as either of the PPMs, a greater $\alpha$ is required to fully encapsulate accuracy of the model's predictions. Even then, the certainty of such motif identification with increasing $\alpha$ decreases.

In \autoref{cnnevaul}, we draw a one to one correspondence between the actual motif indices and that of the predicted motifs by only examining the indices of the 15 highest-scoring 15-mers in both the actual scores and predicted scores. This is not a strong one-to-one correspondence because the number of inserted motifs actually varies randomly from five to fifteen times sample to sample. 
By design, this is a confounding variable 
When $S_\alpha$ is applied on a sample with five inserted motifs, the returned score is predicted to be an underestimate of the model's prediction. This is due to the fact that this function only examines the highest 15-scoring indices for each sample. In the case of five inserted motifs, there would be ten 15-mers identified as high-scoring motifs, when in reality these are random 15-mers in the sequence. Because those scores are more likely to be present throughout a sequence, there will be less similarity between the indices of the predicted 15 highest-scoring 15-mers and that of the actual 15 highest-scoring 15-mers. This will most likely lead to a decrease in $S_\alpha$.

\vspace{12pt}

\end{document}